\documentclass[conference,10pt]{IEEEtran}

\usepackage{cite}
\usepackage{amsmath,amssymb}
\usepackage{graphicx}
\usepackage{textcomp}
\usepackage{xcolor}
\usepackage{bm}
\usepackage{booktabs}

\begin{document}

\title{Power-Dependent Optimal Detuning of Microring Modulators Under Fabrication Variations}

\author{
\IEEEauthorblockN{Abrahim Hamdan and Ishan Thakkar}
\IEEEauthorblockA{Department of Electrical and Computer Engineering, University of Kentucky, Lexington, KY, USA\\
\textit{akha272@uky.edu, igthakkar@uky.edu}}
}

\maketitle

\begin{abstract}
Wafer-scale simulations of 2000 silicon microring modulators show that correcting fabrication-induced resonance errors improves modulation amplitude at low power, but nonlinear effects at higher power make power-dependent resonance detuning targets preferable to perfect alignment.
\end{abstract}
\vspace{5pt}

\section{Introduction}

Silicon microring modulators are attractive for wavelength division multiplexed (WDM) links because they are compact, wavelength selective, and low power \cite{deCea2019,wu2015}. Their main shortcoming is sensitivity to fabrication variations, which shifts each resonance away from the assigned input wavelength and can severely reduce modulation quality.

On 300\,mm wafers, resonance variation can reach $\sigma_\lambda=0.65$\,nm, or about 81\,GHz \cite{parsons2025}. Since this spread is much larger than the linewidth of a typical microring cavity, an as-fabricated device can operate far from its intended point and produce very low optical modulation amplitude (OMA). The usual remedy is to characterize the resonance and tune the device back to zero detuning with the assigned input signal wavelength \cite{wu2015,wang2020}.

That remedy is compelling in the linear regime, but practical transmitters do not always operate there. At milliwatt input optical powers, thermo-optic shift, two-photon absorption (TPA), free-carrier absorption (FCA), and free-carrier dispersion (FCD) alter the hot-cavity operating point and can move the best modulation condition away from the cold-cavity resonance \cite{deCea2019,gironcastro2024}. Prior nonlinear studies of silicon microrings have also shown bistability and self-pulsation arising from the competition between thermal and carrier effects, reinforcing that detuning must be interpreted dynamically rather than as a fixed cold-cavity quantity. 

This paper quantifies that tradeoff with nonlinear time-domain simulation of 2000 microring modulators whose placements are arbitrarily distributed across a 300\,mm wafer. The main result is that zero-detuning correction is highly effective at low input optical power, but at higher input optical power the best target becomes a nonzero detuning. As a result, wafer-scale characterization of fabrication-induced variations in resonance detuning remains valuable, yet the best use of that information is power-aware correction of resonance rather than universal alignment to exact resonance.

\section{Simulation Framework}

We solve five coupled ordinary differential equations (ODEs) per microring. Four come from the nonlinear temporal coupled-mode model in \cite{gironcastro2024} that track the intracavity field, excess free-carrier density $\Delta N$, and temperature rise $\Delta T$. A fifth ODE models the PN-junction voltage as a first-order RC system with $\tau_{RC}\approx10$\,ps \cite{deCea2019}.


Based on the coupled ODE model in \cite{deCea2019,gironcastro2024}, the total normalized detuning can be written as
\begin{equation}
\tilde{\delta\omega} =
\frac{\omega_p - \omega_0}{\gamma_0}
+
\frac{\omega_0}{n_{\mathrm{Si}}\gamma_0}
\left(
\Delta N\,\frac{dn}{dN}
+
\Delta T\,\frac{dn}{dT}
\right)
-
\frac{d\omega_0/dV_{\mathrm{pn}}}{\gamma_0}\,V_{\mathrm{pn}},
\label{eq:detuning}
\end{equation}
where
\(\omega_p\) is the laser (pump) angular frequency,
\(\omega_0\) is the cold-cavity resonance angular frequency,
and \(\gamma_0\) is the cold-cavity amplitude decay rate (half-linewidth) \cite{deCea2019}.
The parameter \(n_{\mathrm{Si}}\) is the nominal refractive index of silicon,
\(\Delta N\) is the excess free-carrier density,
and \(\Delta T\) is the temperature rise relative to ambient.
The quantities \(dn/dN\) and \(dn/dT\) are the sensitivities of the refractive index \(n\) to carrier density and temperature, respectively.
Finally, \(V_{\mathrm{pn}}\) is the voltage across the microring PN junction, and \(d\omega_0/dV_{\mathrm{pn}}\) is the sensitivity of the resonance angular frequency to that voltage \cite{deCea2019}.
The first term in Eq. \eqref{eq:detuning} is the cold-cavity detuning between the laser and the unperturbed resonance,
the second term is the nonlinear hot-cavity shift due to free carriers and heating,
and the third term is the electrically induced shift from modulation.
Eq. ~\eqref{eq:detuning} highlights the central point of this work:
even if fabrication characterization perfectly removes the cold-cavity offset, the effective operating point still depends on optical power, carrier density, and temperature, so the resonance target that maximizes OMA need not be zero under realistic operating conditions.

We integrate the system with fourth-order Runge--Kutta using 1\,ps steps. The optical input is CW, and modulation uses a 127-bit PRBS-7 with $V_{pp}=4$\,V and $V_{\text{bias}}=-2.5$\,V, and other microring modeling parameters for the ODEs taken from \cite{deCea2019,gironcastro2024}. 
Through-port power is averaged per bit, and the normalized OMA is computed as
\((\mu_1 - \mu_0)/P_{in}\), where \(\mu_1\) and \(\mu_0\) are the mean through-port powers
for logical ``1'' and ``0'' bits, respectively, and \(P_{in}\) is the launched input optical power. We also computed insertion loss (IL). 

We simulate 2000 microring cavities. Resonance offsets are drawn from $\mathcal{N}(0,\sigma_\lambda^2)$ with $\sigma_\lambda=0.65$\,nm, consistent with wafer-scale data in \cite{parsons2025}. Q variation is introduced by scaling loss and coupling lifetime with independent $\mathcal{N}(1,0.10^2)$ factors. We compare two cases: \textit{as-fabricated}, with random resonance and Q variation, and \textit{zero-detuning corrected}, where the resonance offset is removed while Q variation is retained.

\section{Results}

\begin{figure}[t]
\centering
\includegraphics[width=\columnwidth]{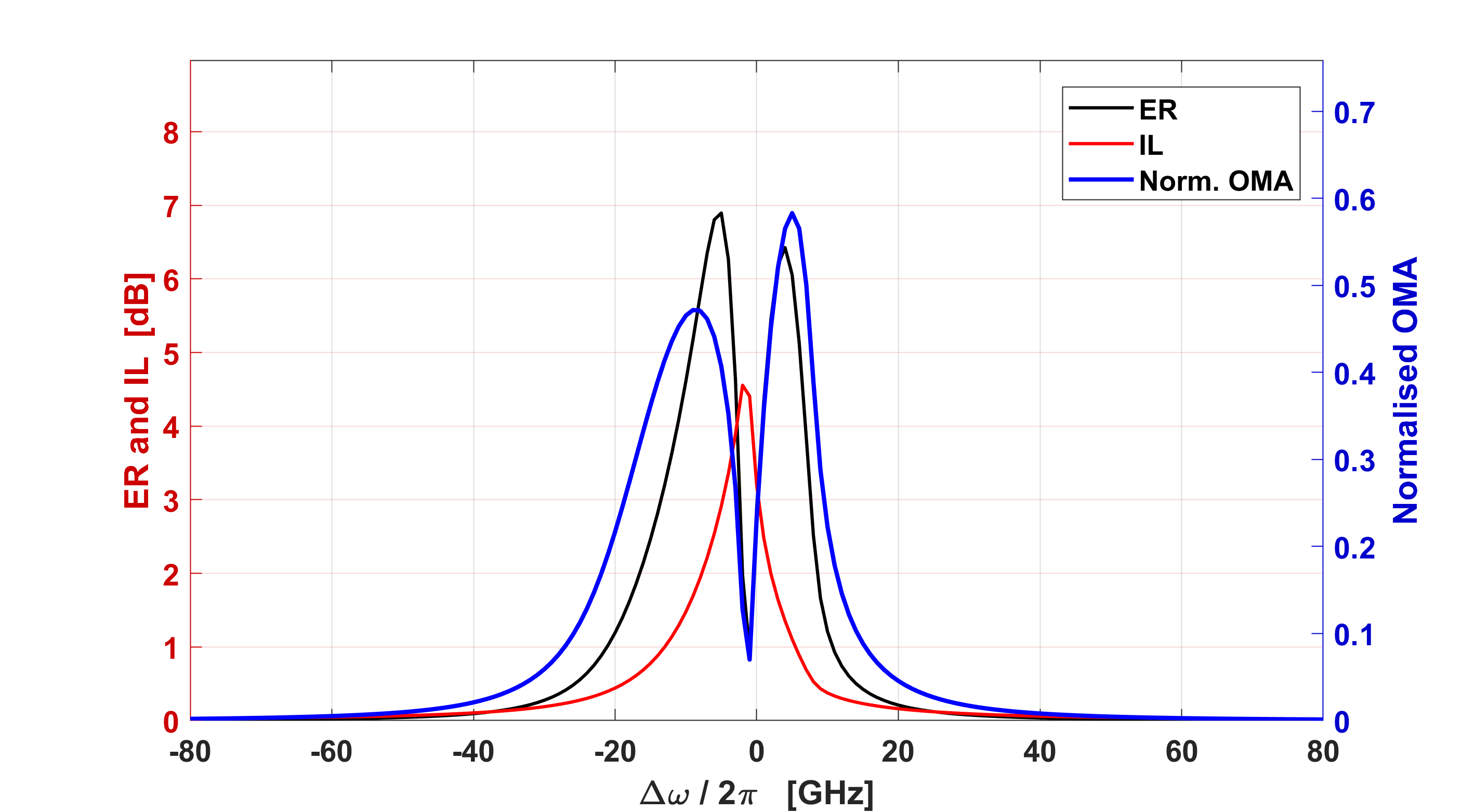}
\caption{ER, IL, and normalized OMA versus detuning at $P_{in}=3$\,mW and 5\,Gbps. OMA peaks near $+10$\,GHz. ER=Extinction Ratio.}
\label{fig:sweep}
\end{figure}

\begin{figure}[t]
\centering
\includegraphics[width=\columnwidth]{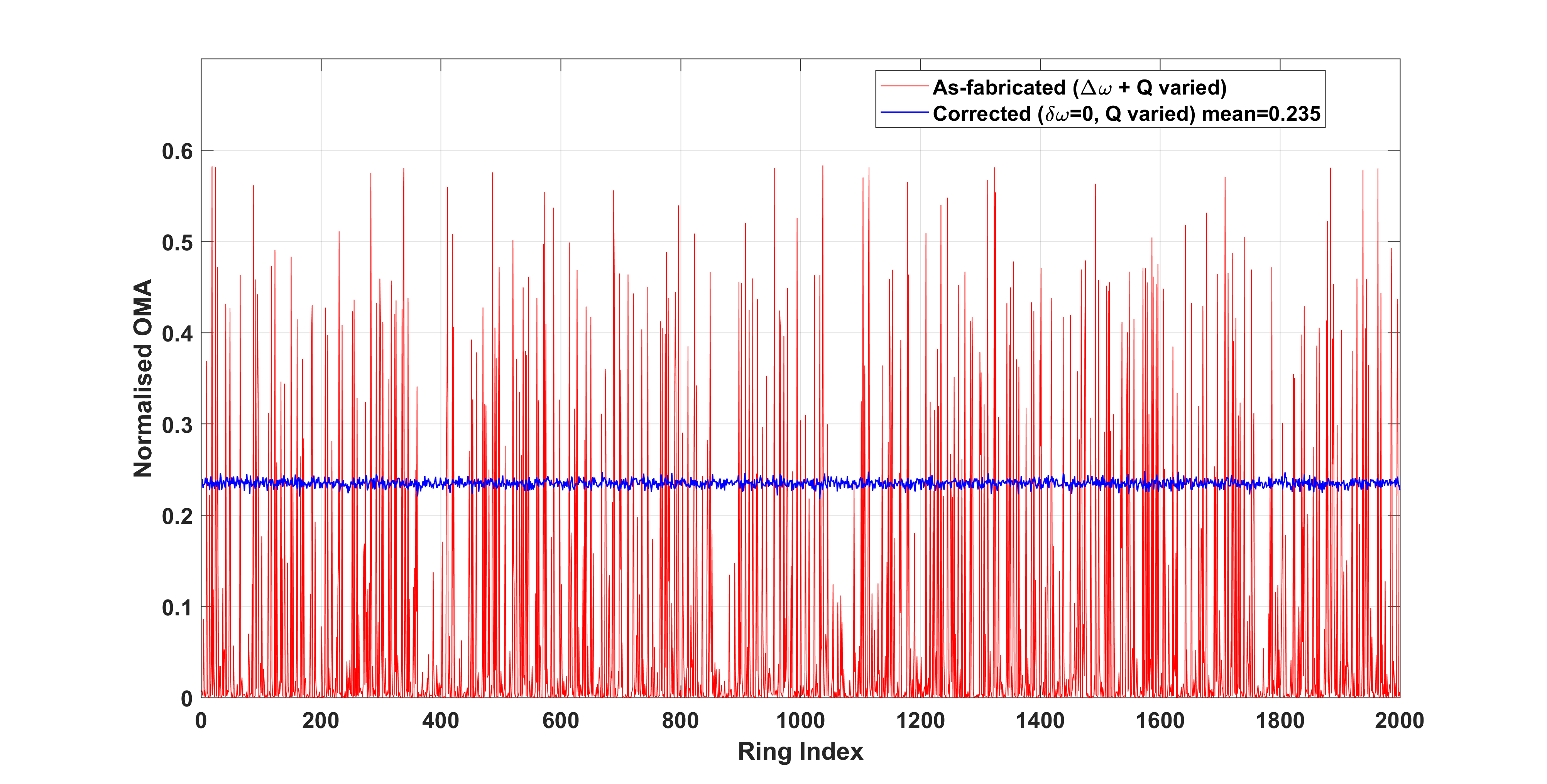}
\caption{Normalized OMA for 2000 microring cavities at $P_{in}=3$\,mW and 5\,Gbps. Red: as-fabricated. Blue: zero-detuning corrected.}
\label{fig:rings}
\vspace{-4pt}
\end{figure}

Fig.~\ref{fig:sweep} shows the central nonlinear effect. At $P_{in}=3$\,mW and 5\,Gbps, the OMA does not maximize at zero detuning. Instead, the OMA peak shifts to about $+10$\,GHz, showing that the best operating point is power dependent \cite{deCea2019}. Physically, this shift arises because TPA-generated carriers and self-heating perturb the resonance in opposite directions, while FCA simultaneously changes the effective loss. The resulting hot-cavity optimum is therefore displaced from the cold-cavity resonance and depends on how much intracavity energy the device stores during modulation \cite{deCea2019}.


Fig.~\ref{fig:rings} shows the ensemble effect. Some as-fabricated microrings already sit near the nonlinear optimum because their detuning happens to be favorable. Forcing zero detuning pushes those rings away from their best operating point, highlighting the core design tension that fabrication characterization reduces detuning uncertainty while a single correction target can become suboptimal in the nonlinear regime.

\begin{table}[t]
\centering
\caption{Mean normalized OMA and IL after zero-detuning correction.}
\label{tab:summary}
\footnotesize
\setlength{\tabcolsep}{3pt}
\begin{tabular}{@{}ccccccc@{}}
\toprule
$P_{\text{in}}$ & Rate & Mean OMA & Mean OMA & Mean IL & Rings w/ & Rings w/ \\
(mW) & (Gbps) & (before) & (after) & after (dB) & better OMA & worse IL \\
\midrule
2 & 2.5  & 0.078 & 0.528 & 1.75 & 96.2\% & 95.5\% \\
2 & 5.0  & 0.076 & 0.480 & 2.08 & 94.4\% & 95.8\% \\
2 & 10.0 & 0.072 & 0.404 & 2.57 & 92.5\% & 96.0\% \\
2 & 15.0 & 0.068 & 0.339 & 2.98 & 90.7\% & 96.5\% \\
\midrule
3 & 2.5  & 0.075 & 0.303 & 2.57 & 87.9\% & 97.2\% \\
3 & 5.0  & 0.073 & 0.235 & 3.21 & 86.0\% & 97.9\% \\
3 & 10.0 & 0.069 & 0.153 & 4.14 & 83.7\% & 98.8\% \\
3 & 15.0 & 0.065 & 0.105 & 4.80 & 81.0\% & 99.0\% \\
\midrule
4 & 2.5  & 0.071 & 0.200 & 4.46 & 84.4\% & 99.5\% \\
4 & 5.0  & 0.068 & 0.188 & 4.44 & 84.5\% & 99.3\% \\
4 & 10.0 & 0.066 & 0.171 & 4.51 & 84.3\% & 98.8\% \\
4 & 15.0 & 0.063 & 0.154 & 4.70 & 83.8\% & 98.8\% \\
\bottomrule
\end{tabular}
\vspace{-4pt}
\end{table}

Table~\ref{tab:summary} shows that zero-detuning correction is strongly beneficial at low power. At $P_{in}=2$\,mW, mean OMA rises from 0.078 to 0.528 at 2.5\,Gbps, and more than 96\% of microring cavities improve. In this regime, nonlinear shifts are modest enough that conventional resonance alignment remains a good approximation.
At higher power, the benefit weakens. At 3--4\,mW, corrected OMA drops and only about 81--88\% of devices improve, while IL worsens in nearly all microrings as tuning toward resonance increases cavity loading. Thus, full correction at high power can produce both higher IL and lower OMA than leaving some detuning. The data-rate dependence is also informative. At 2\,mW, corrected OMA decreases gradually with rate, consistent with finite bandwidth as the bit period approaches electrical and carrier response times \cite{deCea2019}. At 3--4\,mW, the dominant limitation is the unfavorable nonlinear operating point rather than bandwidth, so corrected OMA stays low at all rates.


\section{Conclusion}

Nonlinear simulation of 2000 microring modulators shows that resonance characterization is still valuable, but the best detuning correction target depends on optical power. At low power, tuning to zero resonance detuning strongly recovers modulation amplitude and makes many otherwise weak devices usable. At higher power, thermal and carrier nonlinearities shift the optimum to nonzero detuning, so full correction can reduce modulation amplitude while also increasing insertion loss. Wafer-scale characterization should therefore enable power-aware detuning targets rather than universal alignment to exact resonance, helping decide how much fabrication error to correct for a given operating condition.

\section*{Acknowledgements}
We thank the anonymous reviewers. This work was partially supported by NSF CAREER Award CNS-2450614.


\end{document}